\documentclass[sn-mathphys]{sn-jnl}

\usepackage{makecell}
\usepackage{xcolor}

\jyear{2021}%

\theoremstyle{thmstyleone}%
\theoremstyle{thmstyletwo}%

\theoremstyle{thmstylethree}%

\begin{document}
	
	
	\title[Vibrational resonance in a system of particle on a rotating parabola]{Vibrational resonance in a damped and two-frequency driven system of particle on a rotating parabola}	
	
	\author[1]{\fnm{R} \sur{Kabilan}}\email{kabilanrajagopal82@gmail.com}
	\equalcont{These authors contributed equally to this work.}
	
	\author[1,2]{ \fnm{M} \sur{Sathish Aravindh}}\email{sathisharavindhm@gmail.com}
	\equalcont{These authors contributed equally to this work.}
	
	\author*[1]{\fnm{A} \sur{Venkatesan}}\email{av.phys@gmail.com}
	
	\author[2]{\fnm{M} \sur{Lakshmanan}}\email{lakshman.cnld@gmail.com}
	
	\affil*[1]{\orgdiv{PG \& Research Department of Physics}, \orgname{Nehru Memorial College (Autonomous), Affiliated to Bharathidasan University}, \orgaddress{\street{Puthanampatti}, \city{Tiruchirappalli}, \postcode{621007}, \state{Tamil Nadu}, \country{India}}}
	
	\affil*[2]{\orgdiv{Department of Nonlinear Dynamics, School of Physics}, \orgname{Bharathidasan University}, \orgaddress{\city{Tiruchirappalli}, \postcode{620024}, \state{Tamil Nadu}, \country{India}}}

	
	\abstract{In the present work, we examine the role of nonlinearity in vibrational resonance (VR) of a forced and damped form of a velocity-dependent potential system. Many studies have focused on studying the vibrational resonance in different potentials, like bistable potential, asymmetrically deformed potential, and rough potential. In this connection, velocity-dependent potential systems are very important from a physical point of view (Ex: pion-pion interaction, cyclotrons and other electromagnetic devices influenced by the Lorentz force, magnetrons,  mass spectrometers). They also appear in several mechanical contexts. In this paper, we consider a nonlinear dynamical system with velocity-dependent potential along with additional damping and driven forces, namely a particle moving on a rotating-parabola system, and study the effect of two-frequency forcing with a wide difference in the frequencies. We report that the system exhibits vibrational resonance in a certain range of nonlinear strength. Using the method of separation of motions (MSM), an analytical equation for the slow oscillations of the system is obtained in terms of the parameters of the fast signal. The analytical computations and the numerical studies concur well.}
	
	
	
	
\maketitle

\section{Introduction}
\label{sec1}

Since its observation by Landa and McClintock \cite{landa2000vibrational}, the phenomenon of vibrational resonance (VR) has been studied analytically, computationally, and empirically in several models of nonlinear systems, spanning many fields of science and engineering. Vibrational resonance primarily relies on the cooperation of two input signals with radically different frequency values - a weak low frequency and a fast high frequency - to improve the quality of the response of the weak frequency signal. Theoretical studies and experimental realization of VR have been reported in diverse fields of research.

In particular, the VR phenomenon  has been studied in bistable systems \cite{lakshmanan2003chaos, rajasekar2016nonlinear,baltanas2003experimental}, multistable systems \cite{roy2017vibrational}, ratchet devices \cite{du2016multiple}, excitable systems \cite{deng2014theoretical}, quintic oscillators \cite{guo2020vibrational}, coupled oscillators \cite{sarkar2019vibrational}, overdamped systems, delayed dynamical systems \cite{guo2020vibrational}, asymmetric potential in Duffing oscillators, fractional order potential oscillators \cite{yang2018vibrational}, FHN neural models \cite{ge2020vibrational}, oscillatory networks \cite{qin2018vibrational, baysal2020effects}, biological nonlinear systems \cite{ning2020vibrational}, parametrically excited systems \cite{sarkar2019controlling}, systems with nonlinear damping \cite{laoye2019vibrational}, deformed potentials \cite{vincent2018vibrational} as well as harmonically trapped and roughed potentials \cite{abirami2017vibrational, laoye2019vibrational}. 

More significantly, experimental realizations of VR have been made, particularly in optical system, bistable
surface emitting lasers \cite{chizhevsky2021amplification}, multistable systems, arrays of hard limiters \cite{ren2016theoretical}, and Chua's circuits \cite{ usama2021vibrational}. The potential application of this novel phenomenon has been found in many areas of science, namely energy harvester from bistable oscillator, energy detectors \cite{ren2018generalized}, signal detection, signal transmission and amplification \cite{yao2019inhibitory, jia2020echo}, and the detection of faults in bearings \cite{xiao2019novel}, also in the design of logic gates and memory devices \cite{venkatesh2016vibrational,venkatesh2017implementation,yao2022logical}.

At the vibrational resonance, it is observed that the effective potential is bifurcated. As a result many studies have focused on the effect of potentials. Further the role played by the system parameters, delay terms, fractional order terms, and potential deformation have been also considered on the induction and control of VR \cite{vincent2018vibrational, guo2020vibrational}. 

Despite the fact that VR is realized and observed in many systems, one can realize that they all belong to systems whose potentials are assumed to be polynomial functions. It will be interesting to investigate the possibility and nature of VR in physically interesting nonpolynomial systems. From this consideration, in the present work a mechanical system which corresponds to a nonpolynomial velocity-dependent potential is considered for such a study. 

The model under consideration is a mechanical model with a rich underlying nonlinear dynamics \cite{venkatesan1997nonlinear, mathews1974unique, nayfeh2008nonlinear}. This model often depicts a particle's motion on a rotating parabola. This mechanical model also describes centrifugation equipment, centrifugal forces, industrial events, and a motorbike being driven in a revolving parabolic well in a circus.

As mentioned above, many of the works on VR are concerned with polynomial oscillators and very few works have focused on VR in nonpolynomial oscillators. In particular, VR in driven oscillators with position-dependent mass \cite{roy2021vibrational}, VR in the Morse oscillator \cite{rajasekar2016nonlinear}, VR in a dual-frequency-driven Tietz-Hua quantum well \cite{olusola2020quantum}, identification and parameter estimation of nonpolynomial forms of damping nonlinearity in dynamic systems \cite{chintha2022identification} are few of the works on VR in nonpolynomial oscillators. More detailed investigations on VR in nonpolynomial oscillators are required to unravel the nature of resonance behavior in these oscillators depending upon the various system parameters. In particular, it will be important to explore the role of nonlinearity parameter in contrast to the damping and forcing parameters in the onset of VR phenomenon.

In this connection, there are numerous methods available to approximate the solution of nonlinear equations. Methods such as the Lindstedt–Poincar\'e perturbation method, the multiple scale perturbation method, the harmonic balance method, and the averaging method are quite useful for this purpose. Each one of the above procedures has its own set of advantages and disadvantages, as well as ranges of validity and applicability in specific situations. However, the method of separation of motion (MSM) \cite{roy2021vibrational,vincent2018vibrational} features several significant advantages over the above methods specifically in connection with the study of VR analytically, e.g., the simplicity in application and the transparency of the physical interpretation. Hence, in order to investigate VR theoretically, we used the MSM to divide the equation of motion into a slow motion and a fast motion and obtain a set of integro-differential equations. The composite system is completely solved by this pair of integro-differential equations, which describe the equations of slow oscillations and fast vibrations, respectively, and their superposition. The method of direct separation of motion appears to be a convenient and simple tool for obtaining the equation of a system’s slow motion. It is also applicable to solving many mechanical problems. This MSM method is employed to study vibrational resonance in many nonlinear systems with polynomial potential functions. In the present work, we extend the general framework of the MSM method following the work of Roy-Layinde \textit{et al.} \cite{roy2021vibrational} to study the vibrational resonance in the nonpolynomial oscillator with velocity-dependent potential. Also, in the present work, the role of nonlinearity on VR in a nonpolynomial system of damped, driven particle moving on a rotating parabola is examined. Most of the studies on VR in the literature are concerned with role of damping in either suppressing or increasing the resonance behaviour \cite{rajasekar2016nonlinear}. But in the present system, it is also shown that the VR occurs in an optimal range of nonlinear strength thereby emphasizing the role of nonlinearity. It is quite interesting and entirely different from the VR studies on other polynomial and nonpolynomial oscillators. 

It is an established fact that resonance occurs in a nonlinear oscillator system if the matching of the natural frequency of an underdamped oscillation of the system with an external periodic driving force causes the response amplitude of the system to be enhanced. In recent times, it has been found that the system amplification of response, or the term ``resonance," is being used by adjusting other system parameters, including different kinds of forces.

As a result, many terms are being introduced, namely, stochastic resonance, chaotic resonance, ghost resonance, auto resonance, anti resonance, and so on, depending upon the manifestation of forces or system parameters [details are given in ref.\cite{rajasekar2016nonlinear,vincent2021vibrational}]. 

In the present work, we essentially consider a nonpolynomial nonlinear oscillator model, namely the motion of a particle on a rotating parabola with additional interactions, which is quite different from the generalized Mathews-Lakshmanan oscillator model with additional interactions studied in ref.[32]. The present oscillator is though analogous to a position-dependent mass (PDM) system, the mass is associated with the strength of nonlinearity. In the present paper, we mainly focus on a specific form of vibrational resonance, which is discussed in terms of a slowly driven system's response to variations in the parameters associated with a high-frequency periodic force. In the absence of nonlinearity, the model considered becomes a linear harmonic oscillator. We have shown that no VR occurs in this system without nonlinearity. On the other hand for a low level of nonlinearity maximum response occurs in this model, which is the most novel feature. Thus, it is essential and important to study the effect of nonlinearity in the system. We have shown that the enhancement or suppression of vibrational resonance is possible through variation of the nonlinear system parameter in the presence of low-frequency and high-frequency forces within appropriate parameter regimes.

Our results, while confirming the basic features of VR as identified in the case of PDM Duffing and Morse oscillators, specifically brings out novel features on the effect of nonlinearity. The main contribution of this paper is that the mere presence of nonlinearity alone does not guarantee the emergence of VR, as in the case of polynomial nonlinear oscillators. An appropriate strength of nonlinearity region is solely responsible for the induction and control of VR. Thus, in the present study, we have investigated the role of nonpolynomial nonlinearity in oscillator systems for inducing vibrational resonance. In particular, we have shown that the vibrational resonance is realized and observed within an optimal range of nonlinear strength. From our study, it is inferred that resonance cannot be observed at all parameter values of nonlinearity.

The article is organized as follows : In Section \ref{sec2}, the dynamical model is discussed. Analytical studies and comparison between numerical and analytical studies based on method of separation of motions are given in Sections \ref{sec3} \& \ref{sec4}, respectively.  Finally, the paper is concluded in Section \ref{sec5}.

\section{The dynamical model}
\label{sec2}

In this article, we consider the nonpolynomial oscillator governed by the equation of motion \cite{venkatesan1997nonlinear, mathews1974unique, nayfeh2008nonlinear}

\begin{equation}
(1+\mu x^2)\ddot{x}+\mu x \dot{x}^2+\omega_{0}^2 x =0,
\label{eq1}
\end{equation}
and subjected to two additional periodic forcings and extra linear damping. 

For the analysis of fundamentally nonlinear processes in a non-generic mechanical system that correspond to a well-known model \cite{mathews1974unique,nayfeh2008nonlinear} is Eq.\eqref{eq1} with the Lagrangian, 

\begin{equation}
L= \dfrac{1}{2}\big[ (1+\mu x^2)\dot{x}^2 - \omega_{0}^2 x^2 \big].
\label{eq2}
\end{equation}

\begin{figure}
	\centering
	\includegraphics[width=0.7\linewidth]{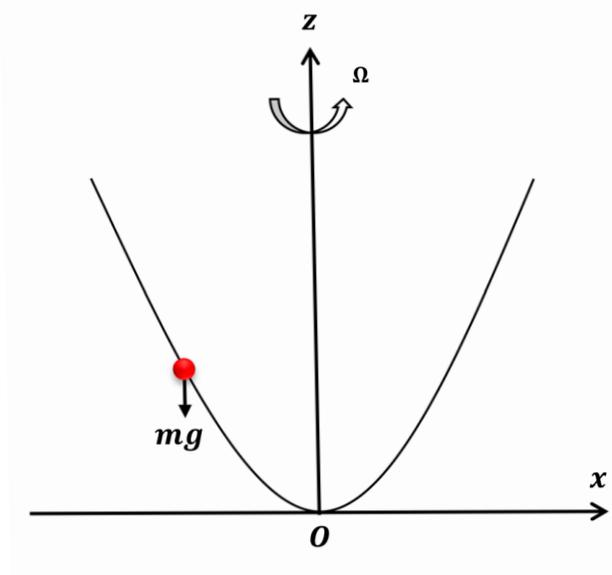}
	\caption{The red circle shows how a freely sliding particle moves down a parabolic wire that rotates at a constant angular velocity $\Omega$ \cite{venkatesan1997nonlinear}.}
	\label{fig1}
\end{figure}

The corresponding Hamiltonian is

\begin{equation}
H = \dfrac{1}{2}\big[p^2 (1+\mu x^2)^{-1} + \omega_{0}^2 x^2 \big],
\label{eq3}
\end{equation}

where the canonical conjugate momentum is

\begin{equation}
p = \dot{x} (1+\mu x^2).
\label{eq4}
\end{equation}

We take into consideration a mechanical model that depicts how a freely moving particle of unit mass moves down a parabolic wire that rotates around the particle's axis $ z=\sqrt{\mu} x^2 $ with a constant angular velocity $\Omega$ ($\Omega^2=\Omega_{0}^2=-\omega_{0}^2+g\sqrt{\mu} $), where $\mu>0$ and $\omega_{0}>0$ as shown in Fig.\ref{fig1}. Here, `g' stands for the acceleration brought on by gravity, `$\frac{1}{\sqrt{\mu}}$' stands for the rotating parabola's semi-lotus rectum, and `$\omega_{0}$' stands for the initial angular velocity \cite{venkatesan1997nonlinear,nayfeh2008nonlinear}.

Next, we investigate the nonlinear dynamics associated with Eq.\eqref{eq1} under the influence of additional damping and two periodic external forcings, leading to the equation of motion  

\begin{equation}
	(1+\mu x^2)\ddot{x}+\mu x \dot{x}^2  + \alpha \dot{x} + \omega_{0}^2 x = f_{1} \cos \omega_{1} t + f_{2} \cos \omega_{2} t.
	\label{eq5}
\end{equation}

In the above $ f_{1} $ and $ f_{2} $ correspond to the strength of two periodic forces with period $ 2\pi/\omega_{1} $ and $ 2\pi/\omega_{2} $ respectively. Here $ \omega_{1} $ and  $ \omega_{2} $ are the strength of the low frequency and high frequency forces, respectively. 

For a single force, Eq.\eqref{eq5} has been well investigated and it shows the period-doubling route to chaos, strange nonchaotic attractors, intermittency, crisis \cite{venkatesan1997nonlinear} and prediction of extreme events using machine learning \cite{meiyazhagan2022prediction}.

We will express Eq.\eqref{eq5} in a way that makes our analytical process manageable, in accordance with the work of Roy-Layinde \textit{et al.} \cite{roy2021vibrational}. In particular, we divide Eq.\eqref{eq5} by $ (1+\mu x^2) $ throughout so that it can be expressed as
\begin{eqnarray}
\ddot{x}+(\mu x \dot{x}^2  + \alpha \dot{x} + \omega_{0}^2 x)(1+\mu x^2)^{-1} \nonumber \\ = (1+\mu x^2)^{-1} (f_{1} \cos \omega_{1} t + f_{2} \cos \omega_{2} t).
\label{eq6}
\end{eqnarray}

Now apply the binomial expansion to the terms $ (1+\mu x^2)^{-1} $ and restrict our consideration to the first three terms of the expansion only. Eq.\eqref{eq6} can be expressed as,
\begin{eqnarray}
\ddot{x}+ (\mu x \dot{x}^2  +  \alpha \dot{x}+ \omega_{0}^2 x)(1-\mu x^2+\mu^2 x^4) \nonumber \\ = (1-\mu x^2+\mu^2 x^4) (f_{1} \cos \omega_{1} t + f_{2} \cos \omega_{2} t).
\label{eq7}
\end{eqnarray}

The potential $ V(x) $ associated with \eqref{eq7} may then be written

\begin{eqnarray}
V(x) = \dfrac{\omega_{0}^2}{2}x^2-\dfrac{\mu \omega_{0}^2}{4} x^4+\dfrac{\mu^2 \omega_{0}^2}{6} x^6.
\label{eq8}
\end{eqnarray}

\begin{figure}[h]
	\centering
	\includegraphics[width=0.7\linewidth]{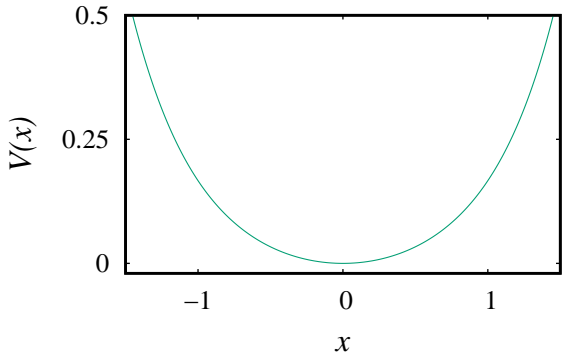}
	\caption{The system potential \eqref{eq8} for $\mu=0.5 $ and $\omega_{0}^2=0.25$.}
	\label{fig2}
\end{figure}

The system potential shown in Fig.\ref{fig2} for given parameter values, $\omega_{0}^2 $ and $ \mu$, is computed from Eq.\eqref{eq8}.

In connection with the analysis of the above nonlinear system, a couple of natural questions arise: In a linear system, is it possible to detect vibrational resonance?  What happens when a linear system is subjected to a high-frequency force? To answer these questions we consider Eq.\eqref{eq5} without the nonlinear term ($\mu=0.0$). The linear equation can be easily integrated and the exact analytical solution is given by

\begin{equation}
	x(t)=A_{1}e^{m_{1}t}+A_{2}e^{m_{2}t}+f_{\omega_{1}} \cos (\omega_{1} t + \phi_{1}) + f_{\omega_{2}} \cos (\omega_{2} t + \phi_{2}),
	\label{eq26}
\end{equation}

where

\begin{equation}
	f_{\omega_{1}} = \frac{f_{1}}{\sqrt{(\omega_{0}^2-\omega_{1}^2)^2+\alpha^2 \omega_{1}^2 }}, 	~~
	f_{\omega_{2}} = \frac{f_{2}}{\sqrt{(\omega_{0}^2-\omega_{2}^2)^2+\alpha^2 \omega_{2}^2 }}
\end{equation}

From the above, it may be observed that the structure of  $ f_{\omega_{1}} $ is independent of the amplitude $ f_{2} $ and the frequency $ \omega_{2} $. Therefore, the amplitude $ f_{\omega_{1}} $ is unaffected by the presence of the high frequency force in a linear system \cite{rajasekar2016nonlinear}. Consequently, no VR occurs in the linear system and so it is important to study the effect of nonlinearity via analytical and numerical methods. These are carried out in the upcoming sections.

\section{Theoretical analysis}
\label{sec3}

The conventional method of separation of motions (MSM), which is a perturbation method, is now used, in which it is assumed that the system dynamics consists of a slow component $ y(t) $ and a fast component $ z(t,\tau) $. The main equation of the system, Eq.\eqref{eq7}, is solved entirely by superimposing the solutions of the two integro-differential equations derived from each of the components using the MSM approach. Following the factorization of like terms and the definition $ x = y + z $, the system equation \eqref{eq7} can be represented as 
	\begin{eqnarray}
		&&\ddot{y}+\ddot{z}+\mu[\mu^2 y^5 + 5 \mu^2 z y^4 + (10 \mu^2 z^2 - \mu)y^3+(10 \mu^2 z^3 - 3 \mu z)y^2+(5\mu^2 z^4 -3 \mu z^2 + 1) y \nonumber \\&& + \mu^2 z^5 - \mu z^3 + z ](\dot{y}^2  +2 \dot{y} \dot{z}  + \dot{z}^2)+\alpha[1+\mu^{2}y^{4}+4\mu^{2}zy^{3}+(6\mu^{2}z^{2}-\mu)y^{2}\nonumber \\&&+(4\mu^{2}z^{3}-2\mu z)y](\dot{y}+\dot{z}) +\omega_{0}^{2}[\mu^{2}y^{5}+5\mu^{2}zy^{4}+(10\mu^{2}z^{2}-\mu)y^{3}+(10\mu^{2}z^{3}-3\mu z)y^{2}\nonumber\\&&+(5\mu^{2}z^{4}-3\mu z^{4}-3\mu z^{2}+1)y+\mu^{2}z^{5}-\mu z^{3}+z]=[1+\mu^{2}y^{4}+4\mu^{2}zy^{3}\nonumber\\&&+(6\mu^{2}z^{2}-\mu)y^{2}+(4\mu^{2}z^{3}-2\mu z)y+\mu^{2}z^{4}-\mu z^{2}](f_{1}cos\omega_{1}t+f_{2}cos\omega_{2}t).
		\label{eq9}
	\end{eqnarray}
Now taking the time average in a period over the fast variable $\tau$ and considering the fact that the fast signal 'z' is rapidly oscillating in the term with period $\dfrac{2\pi}{\omega_{2}}$, we have
	\begin{eqnarray}
		&&\ddot{y}+\mu[\mu^{2}y^{5}+5\mu^{2}\overline{z}y^{4}+(10\mu^{2}\overline{z^{2}}-\mu)y^{3}+(10\mu^{2}\overline{z^{3}}-3\mu\overline{z})y^{2}+(5\mu^{2}\overline{z^{4}}-3\mu\overline{z^{2}}+1)y\nonumber\\&&+\mu^{2}\overline{z^{5}}-\mu\overline{z^{3}}+\overline{z}] (\dot{y}^{2}+\overline{\dot{z}^{2}}) +\alpha[1+\mu^{2}y^{4}+4\mu^{2}\overline{z}y^{3}+(6\mu^{2}\overline{z^{2}}-\mu)y^{2}\nonumber\\&&+(4y^{2}\overline{z^{3}}-2\mu\overline{z})y]\dot{y}+\omega_{0}^{2}[\mu^{2}y^{5}+5\mu^{2}\overline{z}y^{4}+(10\mu^{2}\overline{z^{2}}-\mu)y^{3}+ (10\mu^{2}\overline{z^{3}}-3\mu\overline{z})y^{2}  \nonumber\\&&+(5\mu^{2}\overline{z^{4}}-3\mu\overline{z^{2}}+1)y+\mu^{2}\overline{z^{5}}-\mu\overline{z^{3}}+\overline{z}]=[1+\mu^{2}y^{4}+  4\mu^{2}\overline{z}y^{3}\nonumber\\&&+(6\mu^{2}\overline{z^{2}}-\mu)y^{2}+(4y^{2}\overline{z^{3}}-2\mu\overline{z})y+\mu^{2}\overline{z^{4}}-\mu\overline{z^{2}}+1](f_{1}cos\omega_{1}t+\overline{f_{2}cos\omega_{2}t}).~~~~~
		\label{eq10}
	\end{eqnarray}
In the above, the average value of `z' in relation to the fast variable `$\tau=\omega_{2}t$' is given by \\
\begin{equation}
	\overline{z} =\dfrac{1}{2\pi}\int_{0}^{2\pi}zd\tau=0
	\label{eq11}
\end{equation}
so that Eq.\eqref{eq10} becomes,
	\begin{eqnarray}
		&&\ddot{y}+\mu[\mu^{2}y^{5}+(10\mu^{2}\overline{z^{2}}-\mu)y^{3}+(10\mu^{2}\overline{z^{3}})y^{2}+(5\mu^{2}\overline{z^{4}}-3\mu\overline{z^{2}}+1)y \nonumber\\ &&+\mu^{2}\overline{z^{5}}-\mu\overline{z^{3}}] (\dot{y}^{2}+\overline{\dot{z}^{2}})+\alpha[1+\mu^{2}y^{4}+  (6\mu^{2}\overline{z^{2}}-\mu)y^{2}+4\mu^{2}\overline{z^{3}}y]\dot{y} \nonumber \\ &&+\omega_{0}^{2}[\mu^{2}y^{5}+(10\mu^{2}\overline{z^{2}}-\mu)y^{3}+10\mu^{2}\overline{z^{3}}y^{2}+(5\mu^{2}\overline{z^{4}}-3\mu\overline{z^{2}}+1)y+\mu^{2}\overline{z^{5}}-\mu\overline{z^{3}}] \nonumber \\ 
		&& =1+\mu^{2}y^{4}+(6\mu^{2}\overline{z^{2}}-\mu)y^{2}+(4\mu^{2}\overline{z^{3}}y+\mu^{2}\overline{z^{4}}-\mu\overline{z^{2}}+1)(f_{1}cos\omega_{1}t),~~~~~~
		\label{eq12}
	\end{eqnarray}
Eq.\eqref{eq12} describes the slow motion of the system.

The averages in the slow motion equation can now be found using the approximation method. 
For the composite system '$ x $', this is done by first finding the equation for the initial oscillations in '$ z $' by deducting Eq.\eqref{eq12} from Eq.\eqref{eq9}, which is the equation for the slow component `$ y $'. Consequently, the equation governing the system's fast oscillations can be expressed as 
	\begin{eqnarray}
		&&\ddot{z}+2\mu\dot{y}[\mu^{2}y^{5}+5\mu^{2}zy^{4}+(10\mu^{2}z^{2}-\mu)y^{3}+(10\mu^{2}z^{3}-3\mu z)y^{2}+(5\mu^{2}z^{4}-3\mu z^{2}+1)y \nonumber \\
		&&+\mu^{2}z^{5}-\mu z^{3}+z]\dot{z}+\mu[5\mu^{2}zy^{4}+10\mu^{2}y^{3}(z^{2}-\overline{z^{2}})+10\mu^{2}y^{2}(z^{3}-\overline{z^{3}})-3\mu zy^{2}+\nonumber\\
		&&5\mu^{2}y(z^{4}-\overline{z^{4}})-3\mu y(z^{2}-\overline{z^{2}})+\mu^{2}(z^{5}-\overline{z^{5}})-\mu(z^{3}-\overline{z^{3}})+z]\dot{y}^{2}+[\mu^{2}y^{5}(\dot{z}^{2}-\overline{\dot{z}^{2}})\nonumber\\
		&&+5\mu^{2}z\dot{z}^{2}y^{4}+10\mu^{2}y^{3}(z^{2}\dot{z}^{2}-\overline{z^{2}}\overline{\dot{z}^{2}})-\mu y^{3}(\dot{z}^{2} + \overline{\dot{z}^{2}}) +10\mu^{2}y^{2}(z^{3}\dot{z}^{2}-\overline{z^{3}}\overline{\dot{z}^{2}})-3\mu y^{2}z\dot{z}^{2}\nonumber\\
		&&+5\mu^{2}y(z^{4}\dot{z}^{2}-\overline{z^{4}}\overline{\dot{z}^{2}})-3\mu y(z^{2}\dot{z}^{2}-\overline{z}^{2}\overline{\dot{z}^{2}})+y(\dot{z}^{2}-\overline{\dot{z}^{2}})+\mu^{2}(z^{5}\dot{z}^{2}-\overline{z^{5}}\overline{\dot{z}^{2}})\nonumber\\
		&&-\mu(z^{3}\dot{z}^{2}-\overline{z^{3}}\overline{\dot{z}^{2}})+z\dot{z}^{2}]+\alpha[4\mu^{2}zy^{3}+6\mu^{2}y^{2}(z^{2}-\overline{z^{2}})+4\mu^{2}y(z^{3}-\overline{z^{3}})\nonumber\\
		&&-2\mu zy+\mu^{2}(z^{4}-\overline{z^{4}})-\mu(z^{2} -\overline{z^{2}})]\dot{y}+[\mu^{2}y^{4}+4\mu^{2}zy^{3}+(6\mu^{2}z^{2}-\mu)y^{2}\nonumber\\
		&&+(4\mu^{2}z^{3}-2\mu z)y+\mu^{2}z^{4}-\mu z^{2}+1]\dot{z}+\omega_{0}^{2}[5\mu^{2}zy^{4}+10\mu^{2}y^{3}(z^{2}-\overline{z^{2}})\nonumber\\
		&&+10\mu^{2}y^{2}(z^{3}-\overline{z^{3}})-3\mu y^{2}z+5\mu^{2}y(z^{4}-\overline{z^{4}})-3\mu y(z^{2}-\overline{z^{2}})\nonumber\\
		&&+\mu^{2}(z^{5}-\overline{z^{5}})-\mu(z^{3}-\overline{z^{3}})+z]=[4zy^{3}+b\mu^{2}(z^{2}-\overline{z^{2}})y^{2}+4\mu^{2}(z^{3}-\overline{z^{3}})y-2 \mu zy\nonumber\\
		&&+\mu^{2}(z^{4}-\overline{z^{4}})-\mu(z^{2}-\overline{z^{2}})](f_{1}\cos\omega_{1}t)+[\mu^{2}y^{4}+4zy^{3}+(6z^{2}\mu^{2}-\mu)y^{2}\nonumber\\
		&&+(4\mu^{2}z^{3}-2\mu z)y+\mu^{2}z^{4}-\mu z^{2}+1](f_{2}\cos\omega_{2}t).~~~~~~~
		\label{eq13}
	\end{eqnarray}	

We observe that Eqs.\eqref{eq12} and \eqref{eq13} are a pair of integro-differential equations that, when combined, totally solve the composite system Eq.\eqref{eq7} in an average sense. They describe the equations for the slow oscillations '$ y $' and the fast vibrations $ z $, respectively. Then, using the assumption that the component '$ z $' oscillates considerably faster than the slow component '$ y $', we apply the inertial approximation $\ddot{z}>>\dot{z}>>z>>z^{2}$ and treat the variable '$ y $' as constant in Eq.\eqref{eq13}. Hence Eq.\eqref{eq13} is reduced to\\
\begin{equation}
	\ddot{z} = f_{2} cos\omega_{2}t
	\label{eq14}
\end{equation}
which has a solution
\begin{equation}
	z=-\dfrac{f_{2}}{\omega_{2}^{2}} cos \omega_{2} t,
	\label{eq15}
\end{equation}

leading to the mean values
	\begin{subequations}
		\begin{equation}
				\overline{z}=\overline{z^{3}}=\overline{z^{5}}=0;~~~~~\overline{z^{2}}=\dfrac{f_{2}^{2}}{2\omega_{2}^{4}};		
		\end{equation}
	\begin{equation}		\overline{z^{4}}=\dfrac{3f_{2}^{4}}{8\omega_{2}^{8}};~~~~~\overline{\dot{z}^{2}}=\dfrac{f_{2}^{2}}{2\omega_{2}^{2}}	
	\end{equation}
	\label{eq16}
	\end{subequations}

Using Eq.\eqref{eq16} in Eq.\eqref{eq12}, after simplification it reads as the equation of motion for the slow component as
	\begin{eqnarray}
		&&\ddot{y}+\mu[\mu^{2}y^{5}+(10\mu^{2}\overline{z^{2}}-\mu)y^{3}+(5\mu^{2}\overline{z^{4}}-3\mu\overline{z^{2}}+1)]\dot{y}^{2}+\alpha[\mu^{2}y^{4}+(6\mu^{2}\overline{z^{2}}-\mu)y^{2}\nonumber\\
		&&+\mu^{2}\overline{z^{4}}-\mu\overline{z^{2}}+1]\dot{y}  +[\mu(5\mu^{2}\overline{z^{4}}-3\mu\overline{z^{2}}+1)\overline{\dot{z}^{2}}+\omega_{0}^{2}(5\mu^{2}\overline{z^{4}}-3\mu\overline{z^{2}}+1)]y\nonumber\\
		&&+[\mu(10\mu^{2}\overline{z^{2}}-\mu)\overline{\dot{z^{2}}}+\omega_{0}^{2}(10\mu^{2}\overline{z^{2}}-\mu)]y^{3}  +(\mu^{3}\overline{\dot{z}^{2}}+\omega_{0}^{2}\mu^{2})y^{5}\nonumber\\
		&&=(\mu^{2}y^{4}+(6\mu^{2}\overline{z^{2}}-\mu)y^{2}+\mu^{2}\overline{z^{4}}-\mu\overline{z^{2}}+1)f_{1}\cos\omega_{1}t
		\label{eq17}
	\end{eqnarray}
Keeping in mind the values of various mean values given in Eqs.\eqref{eq16} and redefining the quantities by the following quantities,
\begin{figure}
	\centering
	\includegraphics[width=0.7\linewidth]{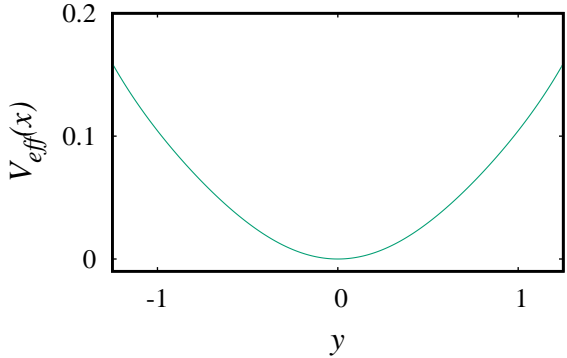}
	\caption{The effective potential \eqref{eq19} for $\mu=0.5 $ and $\omega_{0}=0.25$.}
	\label{fig3}
\end{figure}
\begin{eqnarray}
	C_{1}&=&(5\mu^{2}\overline{z^{4}}-3\mu\overline{z^{2}}+1); \nonumber \\ C_{2}&=&(6\mu^{2}\overline{z^{2}}-\mu); \nonumber \\
	C_{3}&=&(10\mu^{2}\overline{z^{2}}-\mu); \nonumber \\ C_{4}&=&1+\mu^{2}\overline{z^{4}}-\mu\overline{z^{2}},
\end{eqnarray}
\begin{eqnarray}
	\eta_{1}&=&(\mu\overline{\dot{z}^{2}}+\omega_{0}^{2})C_{1}; \nonumber \\ 
	\eta_{2}&=&(\mu\overline{\dot{z}^{2}}+\omega_{0}^{2})C_{3}; \nonumber \\
	\eta_{3}&=&\mu^{2}(\mu\overline{\dot{z}^{2}}+\omega_{0}^{2}),	
	\label{eq24}
\end{eqnarray}
the slow oscillations of the system described by Eq.\eqref{eq12} can be written as
	\begin{eqnarray}
		\ddot{y}+\mu(C_{1}y+C_{3}y^{3}+\mu^{2}y^{5})\dot{y^{2}}+\alpha[C_{4}+C_{2}y^{2}+\mu^{2}y^{4}]\dot{y}+\eta_{1}y+\eta_{2}y^{3}+\eta_{3}y^{5}\nonumber\\
		=(C_{4}+C_{2}y^{2}+\mu^{2}y^{4})f_{1}\cos\omega_{1}t.~~~~~~~~~~
		\label{eq18}
	\end{eqnarray}	
\begin{figure}
	\centering
	\includegraphics[width=0.7\linewidth]{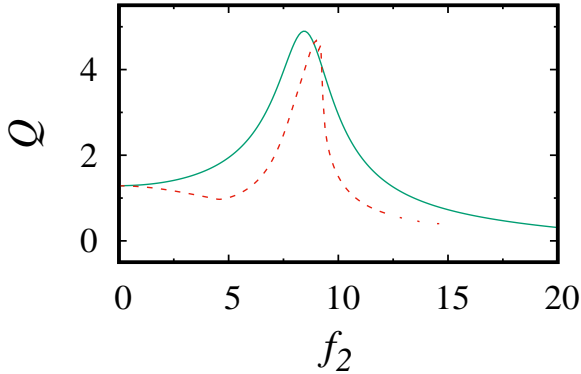}
	\caption{The relationship between $ f_2 $ and the response amplitude $ Q $. The solid line represents analytically computed response amplitude and the dotted line represents the numerical response amplitude for the fixed value of $\mu=0.5$, $ f_{1}=0.05 $ $\alpha=0.2$, $ \omega_{1} = 1.0 $ and $ \omega_{2} = 4.5 $.}
	\label{fig4}
\end{figure}
Consequently, the system's effective potential is given by \\
\begin{equation}
	V_{eff}(y)=\dfrac{\eta_{1}}{2}y^{2}+\dfrac{\eta_{2}}{4}y^{4}+\dfrac{\eta_{3}}{6}y^{6}.
	\label{eq19}
\end{equation}
\begin{figure}[]
	\centering
	\includegraphics[width=0.32\linewidth]{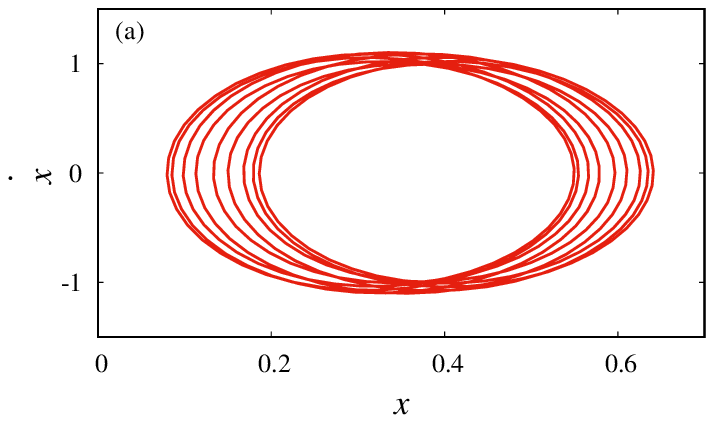}
	\includegraphics[width=0.32\linewidth]{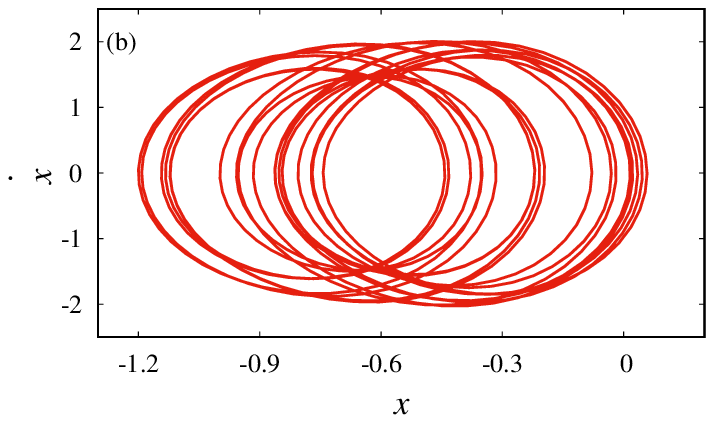}
	\includegraphics[width=0.32\linewidth]{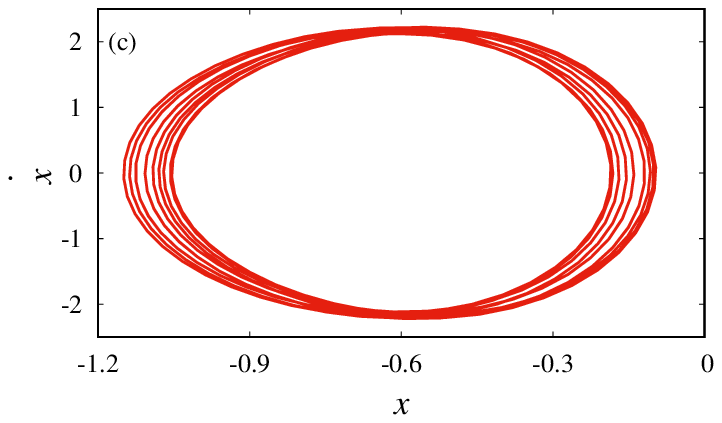}
	\caption{Phase-space trajectories for different values of frequencies of the second periodic force. Panles (a)-(c) represent $ f_{2}=5.0, 9.0 $ and $ 11.0 $, respectively, with fixed value of $\mu=0.5$, $ f_{1}=0.05 $ and $\alpha=0.2$.}
	\label{fig61}
\end{figure}

The linearized equation of motion becomes Eq.\eqref{eq18} by neglecting the nonlinear components and applying the approximation $f_{1}<<1$ such that $\mid{y}\mid<<1$ in the long-term limit $t\rightarrow \infty$: 
\begin{equation}
\ddot{y}+\lambda\dot{y}+\omega_{r}^2 y=F \cos\omega_{1}t,	
\end{equation}
where the resonant frequency is $\omega_{r}=\sqrt{\eta_{1}}$, $\lambda=\alpha C_{4}$ and $F=C_{4}f_{1}$. It is clear that the steady state solution of equation Eq.\eqref{eq18} takes the form $y(t) = A_{L}\cos(\omega_{1}t+\phi) $.
We define the quantity $ A_{L}/f_{1} $ as the response amplitude `$ Q $'.  The oscillation amplitude $ A_{L} = \frac{F}{\sqrt{(\omega_{r}^2-\omega_{1}^2)^2+\lambda^2\omega_{1}^2}}  $, $\phi = \tan^{-1}\bigg(\dfrac{\lambda \omega_{1}}{\omega_{1}^2-\omega_{r}^2}\bigg)$ and the response amplitude can be computed as              
\begin{equation}
	Q_{ana}=\dfrac{A_{L}}{f_{1}}=\dfrac{C_{4}}{\sqrt{(\omega_{r}^2-\omega_{1}^2)^2+\lambda^2\omega_{1}^2}}.
	\label{eq25}
\end{equation}
Using the original definitions of the various quantities occurring in \eqref{eq25}, the response amplitude can be computed as
	\begin{eqnarray}
		Q_{ana}=\dfrac{1+\frac{3\mu^{2}f_{2}^{4}}{8\omega_{2}8}-\frac{\mu f_{2}^{2}}{2\omega_{2}4}}{[{((\omega_{0}^{2}+\frac{\mu f_{2}^{2}}{2\omega_{2}^{2}})(\frac{15\mu^{2}f_{2}^{4}}{8\omega_{2}^{8}}-\frac{3\mu f_{2}^{2}}{2\omega_{2}^{4}}+1)-\omega_{1}^{2}})^{2}+\alpha^{2}({1+\frac{3\mu^{2}f_{2}^{4}}{8\omega_{2}^{8}}-\frac{\mu f_{2}^{2}}{2\omega_{2}^{4}}})^{2}\omega_{1}^{2}]^\frac{1}{2}}.~~~~~~~
		\label{eq20}	
	\end{eqnarray}
Now we analyze the VR using Eq.\eqref{eq20} and we verify numerically the above theoretical results. In order to appreciate the nature of $ Q_{ana} $, we consider the quantity
$ S={{((\omega_{0}^{2}+\frac{\mu f_{2}^{2}}{2\omega_{2}^{2}})(\frac{15\mu^{2}f_{2}^{4}}{8\omega_{2}^{8}}-\frac{3\mu f_{2}^{2}}{2\omega_{2}^{4}}+1)-\omega_{1}^{2}})^{2}} {  +\alpha^{2}({1+\frac{3\mu^{2}f_{2}^{4}}{8\omega_{2}^{8}}-\frac{\mu f_{2}^{2}}{2\omega_{2}^{4}}})^{2}\omega_{1}^{2}} $ in the denominator of equation \eqref{eq20} and we observe that the qualitative feature of $ Q_{ana} $ can be deduced from the nature of S. That is $ Q_{ana} $ reaches the maximum value provided the value of S is minimum. It is clear from the denominator of equation \eqref{eq20}, the appearance of resonance depends on the system parameters $\omega_{0}^2$, $\mu$, $\alpha$, $ f_{2} $, $\omega_{1}$, and $\omega_{2}$. Now we consider the nonlinear parameter and study its effect. We compute numerically by analyzing the cases $ dS/df_{2}=0 $ and $ d^2 S/df_{2}^2>0 $ at resonance. The results are depicted in Fig.\ref{fig4}.
\section{Numerical Analysis}
\label{sec4}
Next, the theoretical response amplitude $ Q_{ana} $ provided by Eq.\eqref{eq20} has been compared with the numerical $ Q_{num} $ obtained from the Fourier spectrum of the solution of the velocity-dependent equation represented as coupled first-order autonomous ordinary differential equations (ODEs) of the following form:

\begin{eqnarray}
\dfrac{dx}{dt} &=& y, \nonumber \\
\dfrac{dy}{dt} &=&( -\mu x \dot{x}^2  - \alpha \dot{x} - \omega_{0}^2 x + f_{1} \cos \omega_{1} t + f_{2} \cos \omega_{2} t ) \times \nonumber \\ &&~~~~~~~~~~~~~~~~~~~~~~~~ (1-\mu x^2+\mu^2 x^4).
\label{eq21}
\end{eqnarray}

To substantiate our analytical study, we numerically integrate Eq.\eqref{eq21} using the Runge-Kutta Fourth order (RK4) method with step size $\Delta t = 0.01 t$ over a simulation of finite time interval $T_{s}=nT$, with $T=2\pi/\omega_{1}$ being the period of oscillations, where the amplitude $ f_{1} $ and frequency $ \omega_{1} $ belong to the slow component and the amplitude $ f_{2} $ and frequency $ \omega_{2} $ are considered as belonging to the fast component, and n=1,2,3... is the number of complete oscillations. For our computation we fix the system parameter values as $\mu=0.5$, $\omega_0^2=0.25$, $\alpha=0.2$, $f_{1}=0.05$ and $\omega_{1}=1.0$. The remaining factors, $f_{2}$ and $\omega_{2}$, are selected in a way that promotes the emergence of VR. 

\begin{figure}[]
	\centering
	\includegraphics[width=0.47\linewidth]{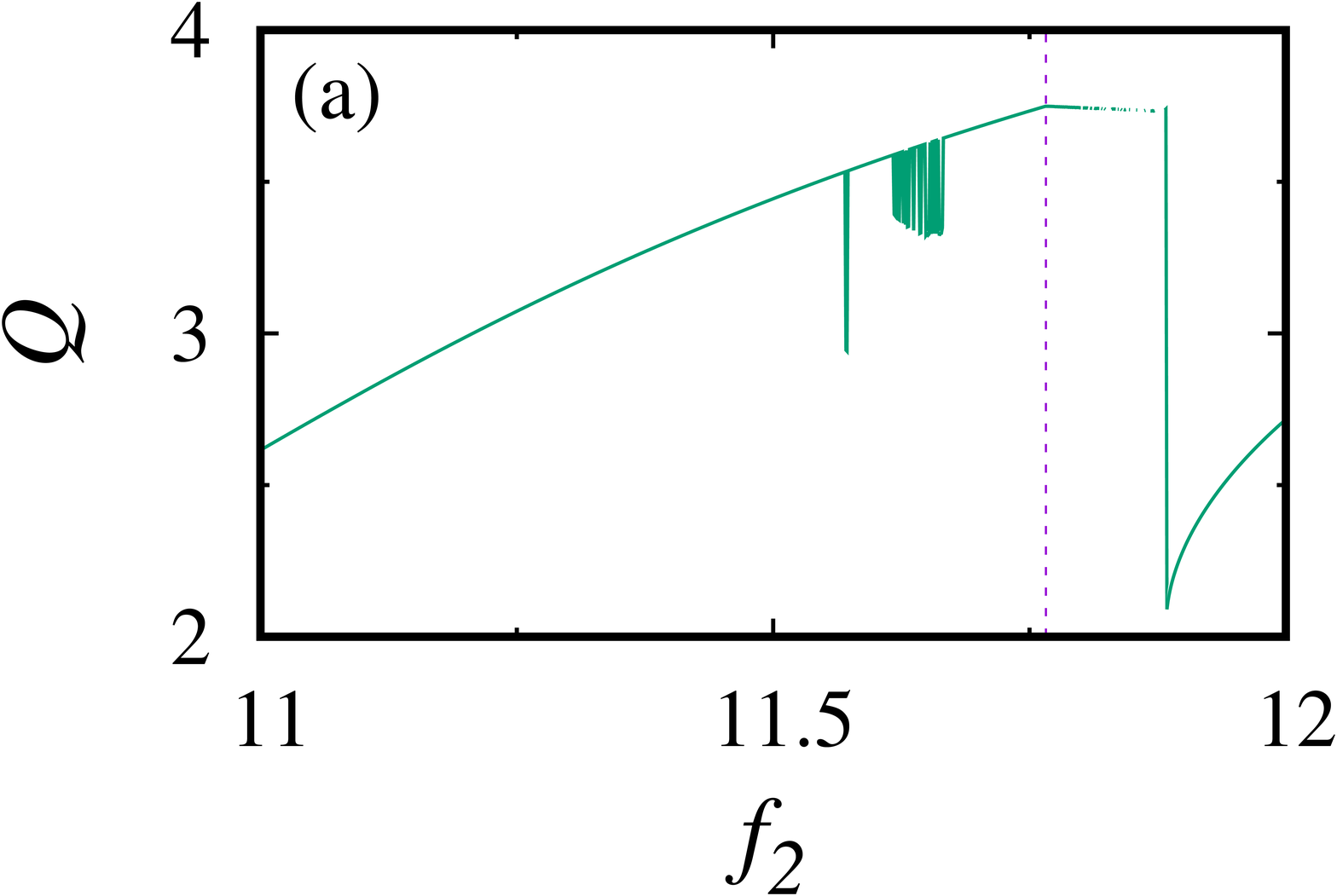}
	\includegraphics[width=0.47\linewidth]{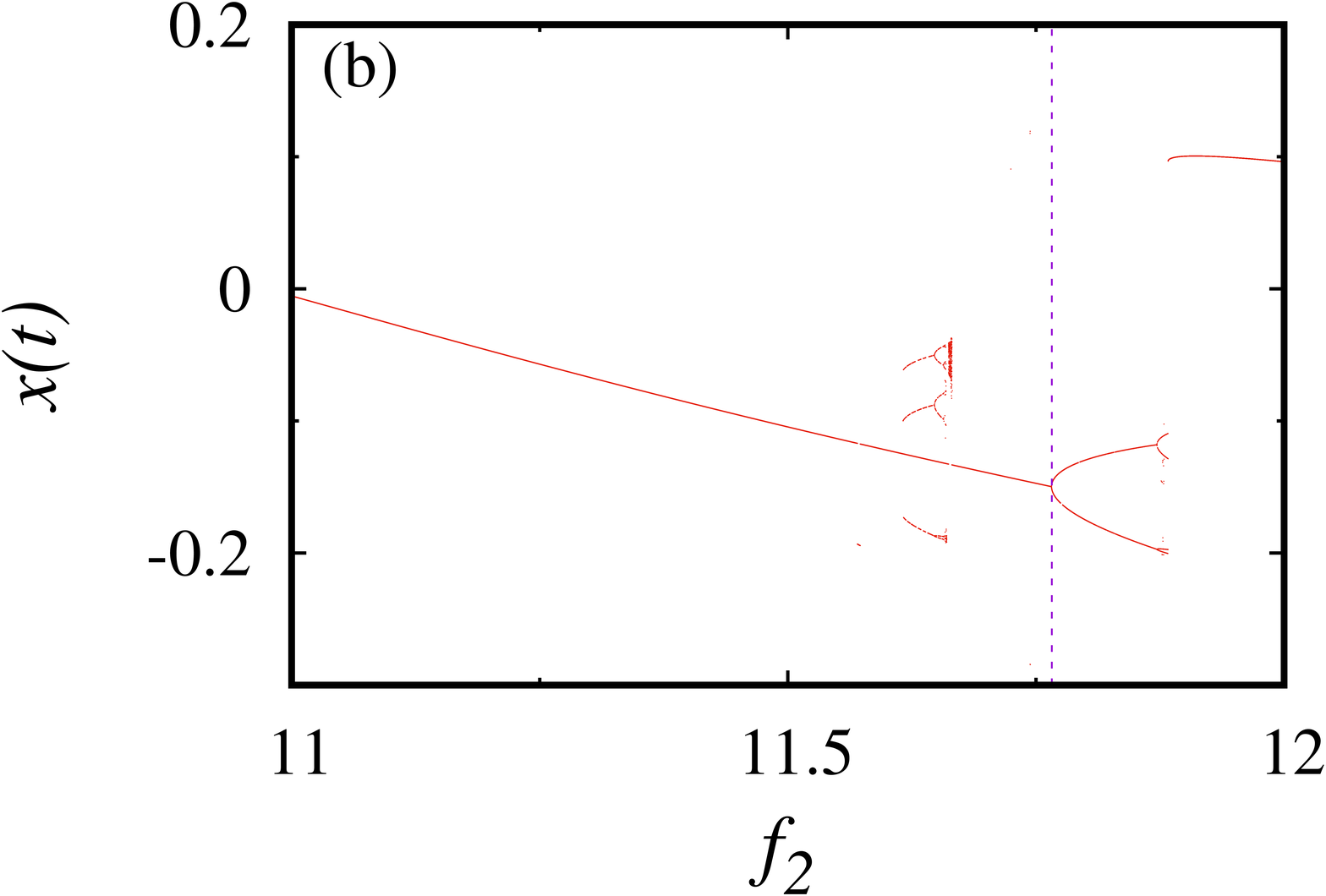}
	\caption{The corresponding response curve Q and the bifurcation diagram of the displacement $ x(t) $ for $\omega_{1}=1.5$. The remaining parameters are fixed as $\mu=0.5$, $ f_{1}=0.05 $ and $\alpha=0.2$.}
	\label{fig6}
\end{figure}

\begin{figure}
	\centering
	\includegraphics[width=0.8\linewidth]{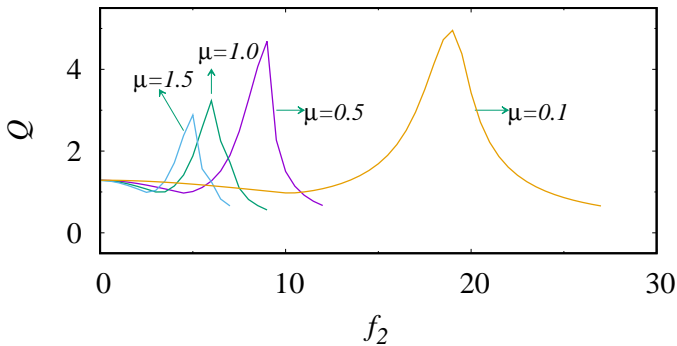}
	\caption{The relationship between $ f_2 $ and the response amplitude $ Q $. The solid lines represent the response amplitudes for different values of $ \mu $ and that curves for $\mu=0.1,0.5,1.0 \text{~and~} 1.5$ are represented by different colors.}
	\label{fig7}
\end{figure}

\begin{figure}[h!]
	\centering
	\includegraphics[width=0.5\linewidth]{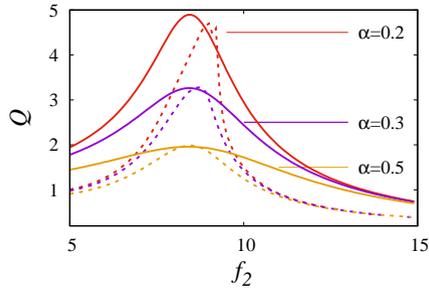}
	\caption{The dependence of response amplitude $ Q $ with $ f_{2} $. The  analytical and numerical results are represented by the solid and dotted line curves, respectively, for different value of $\alpha$.}
	\label{fig10}
\end{figure}

\begin{figure}[h]
	\centering
	\includegraphics[width=0.7\linewidth]{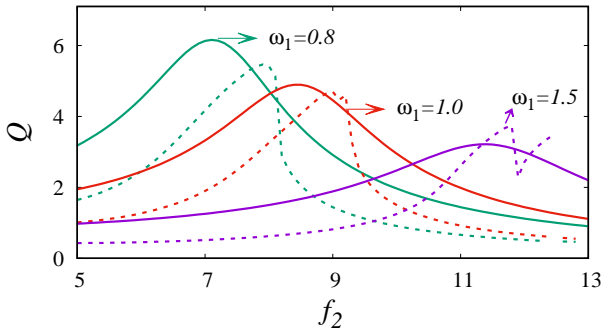}
	\caption{The relationship between $ f_2 $ and the response amplitude $ Q $. The solid line represents analytically computed response amplitude and the dotted line represents the numerical response amplitude for different values of $\omega_{1}$ and that curves for $\omega_{1}=0.8,1.0$ and $ 1.5 $ are represented by different colors.}
	\label{fig8}
\end{figure}
\begin{figure*}[t]
	\centering
	\includegraphics[width=0.47\linewidth]{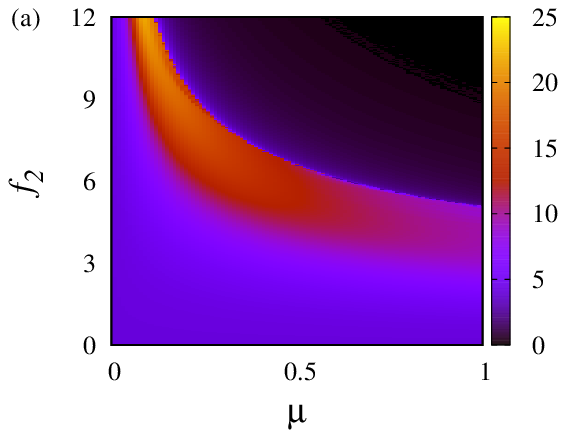}
	\includegraphics[width=0.47\linewidth]{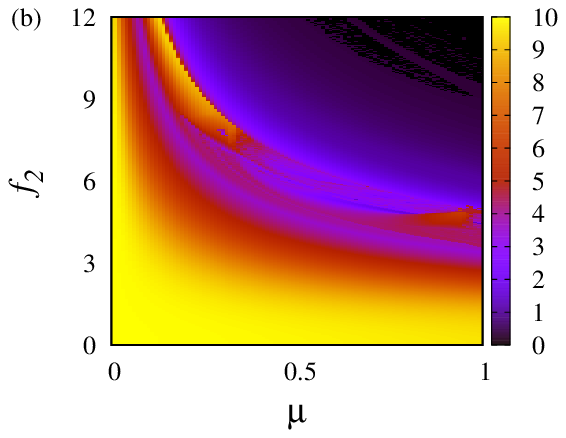}
	\includegraphics[width=0.47\linewidth]{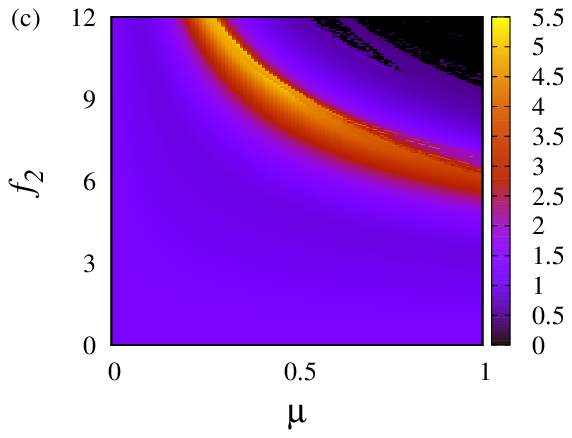}
	\includegraphics[width=0.47\linewidth]{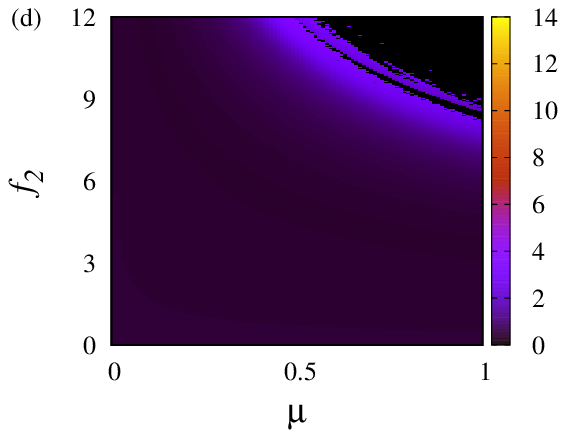}
	\caption{Two-parameter phase diagram for $\mu$ vs second force $ f_{2} $ for fixed $\omega_{1}=0.1,0.5,1.0,1.5$: the maximum response amplitude $ Q $ as indicated in the figure.}
	\label{fig9}
\end{figure*}
After solving numerically Eqs.\eqref{eq21}  and using the results in the following expression for the response function after discarding the transients, we obtain \cite{landa2000vibrational}
\begin{equation}
Q_{num} = \dfrac{\sqrt{\rule{0pt}{2ex} Q_{s}^2+Q_{c}^2}}{f_{1}}, 
\label{eq22}
\end{equation}
over a range of values of the forcing strength of high frequency driving force ($f_2$). Here the values of the quantities $Q_{s}$ and $Q_{c}$ are computed from the Fourier spectrum of the time series of the output signal $x(t)$ as

\begin{eqnarray}
Q_{c} = \dfrac{2}{nT} \int_{0}^{nT} x(t) \cos \omega t dt, \nonumber \\
Q_{s} = \dfrac{2}{nT} \int_{0}^{nT} x(t) \sin \omega t dt.
\label{eq23}
\end{eqnarray}

The response curves of $ Q $ versus selected parameters for a variety of system parameters are superposed to compare the numerically calculated $ Q $ with the previously obtained analytical results. The maximum of each of the analytical curves well matches with the numerically computed results and one can observe that the VR is clearly indicated by an adequate difference between the low and high frequencies.

We begin by examining the occurrence of VR in the response curve of the system as given by Eq.\eqref{eq22} above. We fix the parameters as below: $\mu=0.5$, $\omega_0^2=0.25$, $\alpha=0.2$, $f_{1}=0.05$, $\omega_{1}=1.0$, and $\omega_{2}=4.5$. Fig.\ref{fig4} shows the response amplitude $Q$ of the system depending on the amplitude of high-frequency drive forcing strength $f_{2}$.

It is obvious from Fig.\ref{fig4} that as the fast driving force $ f_{2} $ increases, the response amplitude increases and it reaches a maximum value around $f_{2}=9.0$. Further, as we increase the value of $f_{2}$, the response amplitude $Q$ starts decreasing. The analytically calculated response amplitude from Eq.\eqref{eq20}, shown as a solid line for a set of system parameter values, and Eq.\eqref{eq23}, which is computed directly from the system's main equation, is compared where the corresponding numerical response amplitude shown by broken/dotted lines[see Fig.\ref{fig4} ].

To investigate further, we consider the phase trajectories of the system. These are illustrated in Figs.\ref{fig61}. In Fig.\ref{fig61}(a) the phase space of the trajectories corresponds to a one well structure. Further, on increasing the forcing parameter to the value $ f_{2}=9.0 $, the attractor of the system expands in the phase space and consequently the resonance amplitude grows until it achieves its highest $ Q $ value [see Fig.\ref{fig61}(b)]. On further increase of $f_{2}$ values, the expansion of the attractor decreases and the resonance amplitude also gets decreased [see Fig.\ref{fig61}(c)]. Consequently, it is observed in Fig.\ref{fig4} that the results are consistent with theoretical and numerical studies.

By investigating the underlying dynamics and, in particular, the bifurcation structure, we can now attempt to comprehend the mechanism of VR in the present system. Resonance curves in nonlinear systems are well known to be closely related to the underlying bifurcation structure. In particular, earlier studies support the notion that symmetry-breaking (sb) and  Hopf bifurcations occur between resonances \cite{roy2016analysis,roy2017vibrational}. In this article, we present a new dynamical mechanism linked to resonance. Figs.\ref{fig6}(a-b) show the response curve Q and bifurcation diagram, respectively, obtained by increasing the forcing strength $ f_{2} $. The response curve $ Q $ attains the maximum value just when the period doubling bifurcation takes place at $ f_{2}=11.766 $ on increasing the value of $ f_{2} $. The response curve $ Q $ ﬁrst increases exponentially, and then attains a maximum at the period-doubling  bifurcation point ($ f_{2}=11.766 $). Thus, the maximum in Q signalling VR clearly originates from the period-doubling bifurcation, thereby we confirm that VR is linked to the bifurcation of the attractors.

To investigate how nonlinearity affects the detected vibrational resonance, the response amplitude $ Q $ for different values of $ f_{2} $ on the value of the nonlinearity ($ \mu=0.1,0.5,1.0,1.5 $) of the system is presented in Fig.\ref{fig7}. It is interesting to note that decreasing the strength of the nonlinearity significantly enhances the response amplitude $ Q $ and it reaches a maximum amplitude for $ \mu<0.1 $. Below this value of $\mu$, the Q value remains unchanged. Enhancing the vibrational frequency in the nonlinear system requires an understanding of the role of the damping coefficient $\alpha$. In our study, on varying the damping term $\alpha$ ($\alpha=0.2,0.3$, and $ 0.5 $), our findings demonstrate that the response amplitude $ Q $ diminishes as the damping term $ \alpha $ is increased (see Fig. \ref{fig10}).

To gain more insight on the VR  phenomenon in the present mechanical model, the response amplitude $ Q $ for different `$ \omega_{1} $' values is investigated for various values of the strength of high-frequency force $ f_{2} $. It is observed that on increase in the forcing strength of high-frequency causes a decrease in the response amplitude [see Fig.\ref{fig8}].

Fig.\ref{fig9} is a 3-dimensional plot which depicts the numerically computed response as a function of the nonlinearity $\mu$ and the forcing strength $ f_{2} $ of high frequency for different values of $\omega_1$ ($\omega_1=0.1,0.5,1.0,1.5$). It is clearly demonstrated in Fig.\ref{fig9} that the response amplitude `$ Q $' is high for small values of nonlinearity $\mu$. Increasing nonlinearity value decreases the response amplitude. Further increase in the value of `$\omega_{1}$' also suppresses the response amplitude.

\section{Conclusion}
\label{sec5}

In this paper, we have examined the vibrational resonance (VR) exhibited by a particle moving on a rotating parabola with additional damping force and subjected to combined low-frequency and high-frequency driving forces using analytical and numerical studies. It is found that nonlinearity contributes significantly to the occurrence of VR. The origin and mechanism of vibrational resonance in the present system are identified as corresponding to period doubling bifurcation and it has been confirmed numerically. Usually, in many nonlinear systems, the VR stretches along the nonlinearity, provided the other parameters are appropriately chosen. But in the present model, nonlinearity plays a vital role in the occurrence of VR. It is an established fact that the presence of nonlinearity induces secondary resonances apart from primary resonances. In most of the cases, the nonlinearity is essential for the occurrence of VR. However, the changes in the values of the nonlinear parameter does not make a big difference in the emergence of VR. In the present study, we have shown that the values of the nonlinear parameter in the present nonpolynomial system significantly contributes to the emergence of VR. In particular, we have observed that VR occurs in an optimal range of the nonlinear parameter.

It is established in our study that the induction or control of VR in a nonpolynomial system is due to an appropriate range of a nonlinear system parameter. Our results, while confirming the basic features of VR as identified in the case of PDM Duffing \cite{roy2021vibrational} and Morse oscillators \cite{abirami2017vibrational}, specifically brings out novel features on the effect of nonlinearity.

Also, the nonlinearity is represented by the semi-latus rectum of the rotating parabola and is also part of the angular velocity of the parabola. Thus, the appropriate range of the nonlinear strength in the rotating parabola mechanical system will cause resonance in the system. From a practical point of view, this will lead to the malfunctioning or destruction of parts of the rotating parabola machines, like centrifugation equipment, industrial hoppers, etc. Thus, the present analysis of VR in a particle moving on a rotating parabola is necessary from a technological point of view also. Further, the model is analyzed as a potential-dependent mass (PDM) system, where the mass is defined as the strength of nonlinearity. Thus, the study of nonlinearity in the system reveals the role played by the PDM of the system.

\bmhead{Acknowledgment} 

A.V. expresses his gratitude to the DST-SERB for funding a research project with Grant No.EMR/2017/002813. Sincere appreciation is extended by M.S. to the Council of Scientific \& Industrial Research, India for funding his fellowship via SRF Scheme No.08/711(0001)2K19-EMR-I. A.V. additionally thanks the DST-FIST for funding research projects via Grant No.SR/FST/College-2018-372(C). M.L. thanks the DST-SERB National Science Chair program for funding under Grant No.NSC/2020/000029. 

\subsection*{Author contributions}

RK contributed in setting general idea and implementation its analytical analysis as well as in writing manuscript draft. MS contributed in numerical simulations as well as in writing manuscript draft. AV and ML participated in choosing the methods for problem treatment and presentation of the simulation results as well as writing the paper draft.

\bmhead{Data availability} The datasets generated during and/or analyzed during the current study are available from the corresponding author on reasonable request.



\end{document}